\documentclass{article}

\usepackage{graphicx}
\begin{document}

\title{Cancellation of energy-divergences in Coulomb gauge QCD}

\author{A. Andra\v si  \footnote{aandrasi@rudjer.irb.hr} \\
{\it `Rudjer Bo\v skovi\' c' Institute, Zagreb, Croatia} \\ \\
  \and
J.~C. Taylor\footnote{jct@damtp.cambridge.ac.uk} \\
{\it Department of Applied Mathematics and Theoretical Physics,}\\
{\it University of Cambridge, UK} }

\date{11 April 2005}
\maketitle

\begin{abstract}

{\noindent In the Coulomb gauge of nonabelian gauge theories there are in general,
in individual graphs,
`energy-divergences' on integrating over the loop energy variable
for fixed loop momentum. These divergences are avoided in the Hamiltonian,
phase-space formulation. But, even in this formulation, energy-divergences
re-appear at 2-loop order. We show in an  example how these cancel between
graphs as a consequence of Ward identities.}\\

\noindent{Pacs numbers: 11.15.Bt; 11.10.Gh} \\

\end{abstract}

\vfill\newpage

\section{Introduction}

In gauge theories, the Coulomb gauge has special position.
The number of dynamical variables is the same as the number of physical degrees of
freedom. Moreover, if we go to the Hamiltonian, phase-space, first-order
formalism, there are no ghosts; and, because of the existence of a Hamiltonian,
unitarity should be manifest.

Nevertheless, there are complications, if not problems, with the Coulomb
gauge. In the Lagrangian, second-order, formalism, there are 
`energy-divergences' in individual Feynman graphs. These are
divergences over the energy-integration, $\int dp_0$, in a loop,
for fixed values of the 3-momentum ${\bf p}$. These are difficult to
regularize: dimensional regularization does not touch them, and any other
regularization risks doing violence to gauge-invariance (but see Leibbrandt et al for
a modified form of dim. reg. \cite{leibbrandt}).
These energy-divergences do cancel when all graphs are combined
\cite{mohapatra}; but it makes one uneasy to be manipulating
divergent and unregulated integrals.

The problem of energy-divergences is eased by going to
the Hamiltonian, phase-space, first-order formalism, in which
time derivatives of the gluon field ${\bf A}$ are eliminated in favour
of the conjugate momentum field ${\bf E}$. This has the advantage
of being a true Hamiltonian formalism, and unitarity should be
manifestly obeyed. Also, there are no ghosts (ghost loops cancel
part of the closed Coulomb loops). For a sample calculation in this
formalism see \cite{andrasi}, and for possible connection to confinement
see \cite{CZ}.

But there are still problems. There is a question of operator-ordering
in the Hamiltonian \cite{christlee} (see also \cite{schwinger},  \cite{cheng-tsai}),
which may require higher-order terms.
It has been shown \cite{doust,taylor}  that these
operator-ordering problems are connected with ambiguous multiple
energy-integrals in higher orders.

In this paper, we are concerned with a simpler problem, which arises
at 2-loop order. In general, in the Hamiltonian formalism to this order, the two
integrals over the internal energies converge with the two internal spatial momenta
held fixed. However, renormalization demands that we first perform the energy and
momentum integrals for each subgraph, then make subtractions for ultraviolet divergences,
and then perform the remaining energy and momentum integrals. With this sequence of
operations, one does in general find an energy-divergence in the final energy integral.

 We illustrate this with a simple example in which quark-loop subgraphs are inserted
into the second-order gluon self-energy graphs. We perform the energy-momentum integral in the
subgraph first, then do the renormalization subtraction.
Individual graphs now have energy-divergences in the final energy integral,
but these cancel when graphs are combined.
We show that the cancellation is a consequence of the Ward identities
obeyed by the quark-loop sub-diagrams.

Of course, it is reassuring to check that the energy-divergences do cancel.
But we are back in the uncomfortable position of having to handle
divergent (and unregularized) integrals at intermediate stages of the
cancellation. This contrasts with the Feynman gauge, where all integrals
to all orders are unambiguously regularized by dimensional regularization.

\section{Notation, conventions and Feynman rules}

\def\m{{\mu}}
\def\n{{\nu}}
\def\l{{\lambda}}
\def\s{{\sigma}}
\def\P{{\bf p}}
\def\K{{\bf k}}
\def\Q{{\bf q}}
\def\P'{{bf p'}}

Lorentz indices are denoted by Greek letters, spatial indices by $i,j,...$,
and colour indices by $a,b,c,...$.
We use the metric tensor $g_{\m\n}$, where
\begin{equation}
g_{00}=1,~~~~g_{ij}=-\delta_{ij}.
\end{equation}
Lorentz vectors are written
\begin{equation}
p=(p_0;{\bf p}),~~~ p^2=p_0^2-{\bf p}^2.
\end{equation}
We define the Lorentz tensor $G_{\m\n}$ by
\begin{equation}
G_{ij}=g_{ij},~~~ G_{0\m}=0,
\end{equation}
and for any vector $p_{\m}$ we define the vector $P_{\m}$
by
\begin{equation}
P_i=p_i,~~~P_0=0, ~~~P^2=-{\bf p}^2.
\end{equation}
(Of course, these definitions refer to the particular time-like vector 
$(1;0,0,0)$ with respect to which we have chosen to define the Coulomb
gauge.)
We define a spatial transverse tensor $T$ by
\begin{equation}
T_{\m\n}(p)=-G_{\m\n}+{P_\m P_\n \over P^2},
\end{equation}
so that
\begin{equation}
T_{ij}(p)=\delta_{ij}-{p_ip_j \over {\bf p}^2},~~~T_{\m 0}=0.
\end{equation}

In terms of colour matrices $\tau_a$, $C_q$ is defined by
\begin{equation}
\hbox{tr}(\tau_a \tau_b)=C_q \delta_{ab},
\end{equation}
and in terms of the structure constants $f_{abc}$, $C_G$
is defined by
\begin{equation}
f_{abc}f_{abc'}=C_G \delta_{cc'}.
\end{equation}
The renormalized coupling constant is $g$. Quarks have mass $m$.

We use dimensional regularization with spacetime dimension $4-2\epsilon$.
The Lagrangian density in the phase-space formalism is
\begin{equation}
L=-{1\over 4}(F^a_{ij})^2 + {1\over 2}(E^a_i)^2 - E^a_iF^a_{0i}
\end{equation}
where
\begin{equation}
E^a_iF^a_{0i} = E^a_i[\partial_0A^a_i - \partial_iA^a_0 -gf^{abc}A^b_0A^c_i].
\end{equation}

The Hamiltonian form of the Coulomb gauge has dynamical, conjugate
fields ${\bf A}, {\bf E}$. The Coulomb potential $A_0$ is not a dynamical
variable. It could be eliminated, but we find it convenient to leave
it in the Feynman rules.
We use a continuous line to represent ${\bf E}$, a dashed line for ${\bf A}$
and a dotted line for $A_0$. The propagators are not diagonal in ${\bf A}, {\bf E},A_0$. The propagators are shown in Fig.1 (the arrows on the lines show the direction of the momentum $k$).
Only one vertex will be relevant to our calculation, that shown in Fig.2.

\begin{figure}
\begin{center}
\includegraphics[width=6.2cm]{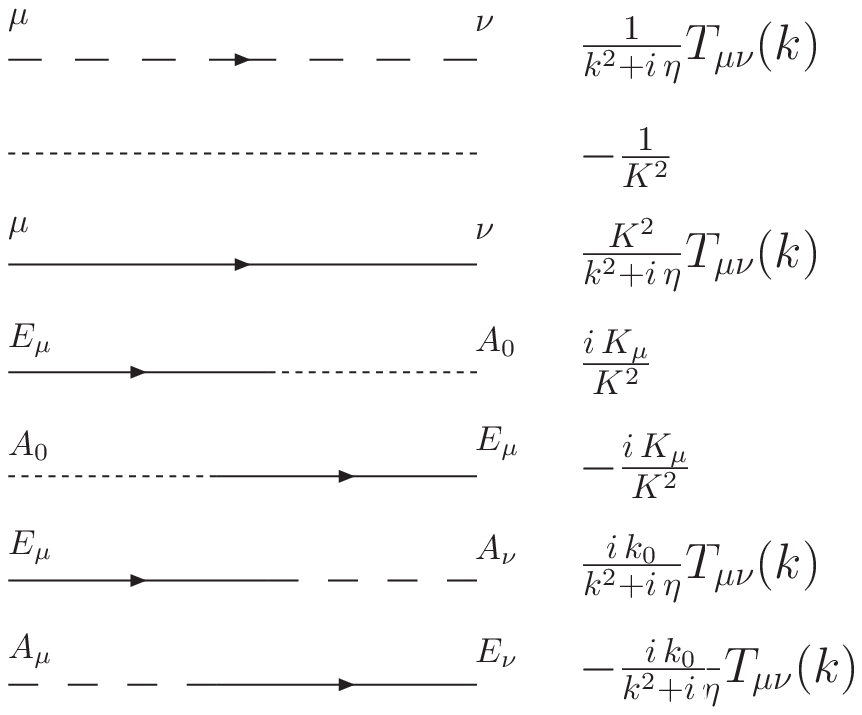}%
\caption{The Feynman rules for propagators in the Hamiltonian Coulomb gauge. Continous lines
represent ${\bf E } $, dashed lines ${\bf A} $ and dotted lines $ A_0 $. The arrow indicates
the sense of momentum flow.}
\end{center}
\end{figure}

\begin{figure}
\begin{center}
\includegraphics[width=5.5cm]{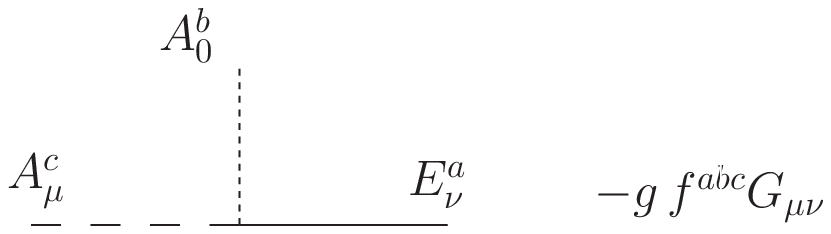}%
\caption{A vertex in the Hamiltonian Coulomb gauge.}
\end{center}
\end{figure}

Its value is
\begin{equation}
-g f_{abc}G_{\m \n}.
\end{equation}
In the Hamiltonian formalism, this is the only vertex involving the Coulomb
field. It is this feature which implies there are no energy-divergences to 
1-loop order.

\section{The quark loop effective action and Ward identities}

We are going to insert quark loops into a gluon diagram, so we need
a notation for quark 1-loop effective action.
Let the 2-gluon term in this effective action be (in momentum space)
\begin{equation}
\delta_{ab}Q_{\m\n}(p)=\delta_{ab}(p^2 g_{\m\n}-p_\m p_\n)Q(p^2){(\mu^2)}^{-\epsilon}.
\end{equation}
We will require to know $ Q $ only for $ ~~ |p_0|\gg |{\bf p}|,~m^2.$ In this region, we have
\begin{equation}
Q(p^2)\sim 8ig^2C_q\pi^{2-\epsilon}\Gamma(\epsilon){{\Gamma^2(2-\epsilon)}\over{\Gamma(4-2\epsilon)}} 
[(-p_0^2-i\eta)^{-\epsilon}-(\mu^2)^{-\epsilon}], 
\end{equation}
(using minimal subtraction with a mass unit $\mu$).

The 3-gluon term in the effective action will be denoted by
\begin{equation}
if_{abc}\Gamma_{\m\n\l}(p,q,p'),
\end{equation}
where $p+q+p'=0$, and the quantum numbers of the three gluons are
$p,\m,a;~~q,\n,b;~~p',\l,c$ (all momenta are directed into the vertex).
We will not need to know the value of $ \Gamma $ in general. 
Finally the 4-gluon term in the effective action will be denoted as
\begin{equation}
W^{abcd}_{\m\n,\l\s}(p,q;k,r),
\end{equation}
where the quantum numbers are $p,\m,a;~~q,\n,b;~~k,\l,c;~~r,\s,d$.
Again, we do not need to know the value of $W$ in general. 
Both $\Gamma$ and $W$ have the symmetries required for Bose symmetry.

Note that the effective action due to quark loops is a functional of
$A_{\m}$, and does not depend upon ${\bf E}$. There are terms in the
effective action which involve the Coulomb field $A_0$, and this is the reason
that energy-divergences re-appear. Renormalization requires also the
presence of counter-terms with a different structure from the
interactions in the original formalism.

The effective action obeys the following Ward identities:
\begin{equation}
p^{\m}\Gamma_{\m\n\l}(p,q,p')=g[Q_{\n\l}(p')-Q_{\n\l}(q)],
\end{equation}
$$ p^{\m}W_{\m\n\l\s}(p,q;k,r)=-gf_{abe}f_{ecd}\Gamma_{\n\l\s}(p+q,k,r) $$
\begin{equation}
-gf_{ace}f_{edb}\Gamma_{\l\s\n}(p+k,q,r)-gf_{ade}f_{ebc}\Gamma_{\s\n\l}(p+r,q,k).
\end{equation} 
These identities express the gauge invariance of the quark loop contribution to the 
effective action. They are a special case of the BRS identities (see for example
Itzykson and Zuber equation (12-144)) when there are no ghost contributions \cite{itzykson}.
We shall show that these identities are sufficient to ensure
the cancellation of energy-divergences between graphs.

\section{The energy-divergent graphs}
The simplest example of the energy-divergences occurs in the gluon 2-point
function, to 2-loop order. The relevant graphs are shown in Fig.3,
where the thick circles represent terms from the quark-loop effective
action (in (12), (14) and (15)). In Fig.3(vi), the sum of the three subgraphs
shown corresponds to (15).

The contributions have the form (in the notation of (4))
\begin{equation}
ig^2\delta_{ab}\int d^{4-2\epsilon}p{1\over {P}^2}[J^{(i)}_{ij}+J^{(ii)}_{ij}+
J^{(iii)}_{ij}+J^{(iv)}_{ij}+J^{(v)}_{ij}+J^{(vi)}_{ij}],
\end{equation}
where the roman numbers correspond to the labels on the diagrams.

\begin{figure}[h]
\begin{center}
\includegraphics[width=9cm]{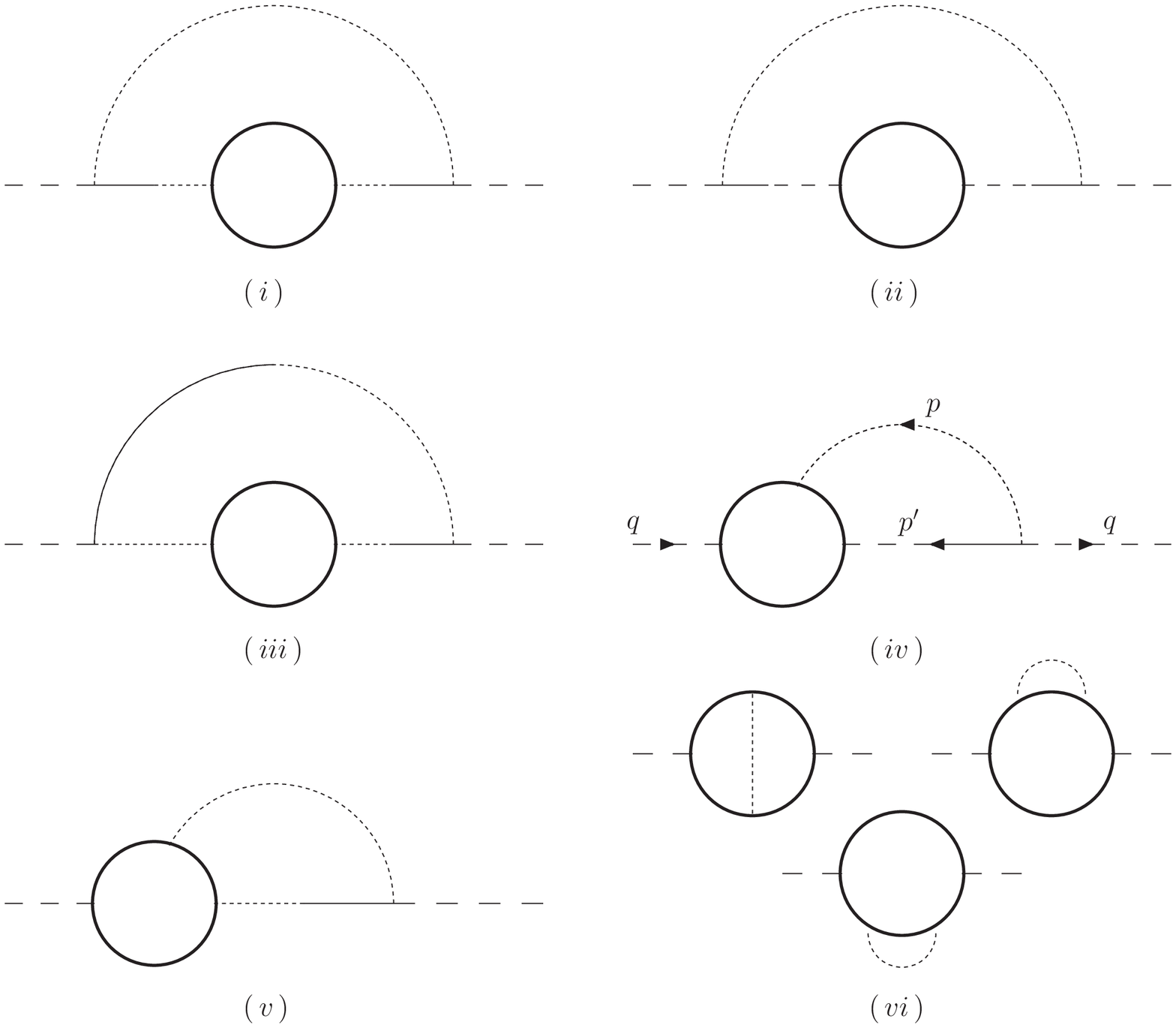}%
\caption{The graphs with energy divergences. The thick black circles represent contributions
from quark loops. The other lines are defined in Fig.2. }
\end{center}
\end{figure}

We have
\begin{equation}
J^{(i)}_{ij}=C_G{P'_i\over P'^2}Q_{00}(p'){P'_j\over P'^2},
\end{equation}
\begin{equation}
J^{(ii)}_{ij}=C_G{p'_0\over p'^2+i\eta}T_{i\m}(p')Q^{\m\n}(p')T_{\n j}(p'){p'_0\over p'^2+i\eta},
\end{equation}
\begin{equation}
J^{(iii)}_{ij}=-C_G{1\over P'^2}P_iQ_{00}(p'){P'_j\over P'^2}+ (i\leftrightarrow j),
\end{equation}
\begin{equation}
J^{(iv)}_{ij}=C_G\Gamma_{i\m 0}(q,p',p){p'_0 \over p'^2+i\eta}T^{\m}_j(p')+(i\leftrightarrow j),
\end{equation}
\begin{equation}
J^{(v)}_{ij}=C_G\Gamma_{i00}(q,p',p){P'_j\over P'^2}+(i\leftrightarrow j),
\end{equation}
\begin{equation}
J^{(vi)}_{ij}={1\over 2}W_{00,ij}(p,q).
\end{equation}
The energy-divergences come from the region of integration where
\begin{equation}
p_0 \gg |{\bf p}|,~q_0,~ |{\bf q}|,~ m. \end{equation}
To examine these divergences, we may use in (19), (20) and (21) 

\begin{equation}
Q_{00}(p')\sim P'^2 Q(p_0),~~~Q_{ij}\sim G_{ij}p_0^2Q(p_0).
\end{equation} 
We then see that (19), (20) and (21) are each  divergent
as integrals over $p_0$ for fixed ${\bf p}$. (Actually, for $\epsilon >0$
this is true only of the contributon from the subtraction term in (13).)
If we take the limit $\epsilon \rightarrow 0$ first, then we get a 
double log energy-divergence.

To find the behaviour of the integrals in (22), (23) and (24),
it is sufficient to use the Ward identities (16) and (17) in the large
$p_0$ limit. Then (16) gives
\begin{equation}
p_0\Gamma_{0ij}(p,q,p') \sim gQ_{ij}(p) \sim gG_{ij}(p)p_0^2 Q(p_0).
\end{equation}
Also,
$$
p_0^2\Gamma_{00i}(p,q,p')\sim p^{\m}p'^{\n}\Gamma_{\m\n i}-p^{\m}P'^{j}\Gamma_{\m ji}
-p'^{\n}P^j\Gamma_{j\n i}$$
\begin{equation}
= -g(P^j-P'^j) G_{ij}p_0^2Q(p_0)=g(P_i-P'_i)p_0^2Q(p_0).
\end{equation}
Similarly, (17) implies that
$$p_0^2 W^{aacd}_{00,ij}(p,-p,q,-q)\sim p_0gC_G\delta_{cd}[\Gamma_{j0i}(p,-p,0)-\Gamma_{ij0}(p,0,-p)] $$
\begin{equation}
\sim 2g^2C_G G_{ij}(p)p_0^2 Q(p_0)\delta_{cd},
\end{equation}
using (27) again.

From (27), (28) and (29), we see that all the integrals in (18) have energy-divergences
of the same kind, and the divergent part has the form
\begin{equation}
ig^2C_G\delta_{ab}\int d^{3-2\epsilon}{\bf p}{1\over{P^2}} dp_0 Q_(p_0)[K^{(i)}_{ij}+...
+K^{(vi)}_{ij}],
\end{equation}
where 
\begin{equation}\label{eq:ovaKi}
K^{(i)}_{ij}= {P'_iP'_j\over P'^2},
\end{equation}
\begin{equation}
K^{(ii)}_{ij}= -T_{ij}(P'),
\end{equation}
\begin{equation}
K^{(iii)}_{ij}= -{P_i P'_j+P_j P'_i  \over P'^2},
\end{equation}
\begin{equation}
K^{(iv)}_{ij}= 2T_{ij}(P'),
\end {equation}\begin{equation}
K^{(v)}_{ij}= {(P_i P'_j+P_j P'_i) \over P'^2} -2{{P'_iP'_j}\over{P'^2}},
\end{equation}
\begin{equation}
K^{(vi)}_{ij}= G_{ij}(P').
\end{equation}
These last six expressions cancel, so the energy-divergences 
in the separate terms in (18) cancel out in the sum.
 
Probably similar cancellations occur
in two-loop graphs made entirely of gluon lines. But in this case there is the extra
complication of the ambiguous integrals connected to the Christ-Lee terms \cite{doust}
\cite{taylor}. \\

Acknowledgement.
This work was supported by MZOS of the Republic of Croatia under Contract No. 0098003 (A.A.).
We are grateful to Dr. G. Duplan\v ci\'c for drawing the figures.

\end{document}